\pdfoutput=1
\documentclass[preprint,12pt]{elsarticle}

\usepackage[T1]{fontenc}
\usepackage[utf8]{inputenc}
\usepackage{microtype}
\usepackage{amsmath,amssymb,mathtools,bm}
\usepackage{graphicx}
\usepackage{hyperref}
\usepackage[nameinlink,capitalise]{cleveref}

\hypersetup{
  colorlinks=true,
  linkcolor=blue,
  citecolor=blue,
  urlcolor=blue,
  pdftitle={Complex chromoelectric polarizability of a heavy-quarkonium resonance: pole response and channel-resolved effective-theory representation},
  pdfauthor={Arkadiy I. Syamtomov}
}

\newcommand{\dd}{\mathrm d}
\newcommand{\ii}{\mathrm i}

\newcommand{\pNRQCD}{\ensuremath{\mathrm{pNRQCD}}}
\newcommand{\cE}{\bm{\mathcal E}}
\newcommand{\ket}[1]{\lvert #1\rangle}
\newcommand{\bra}[1]{\langle #1\rvert}

\journal{Annals of Physics}

\begin{document}
\begin{frontmatter}

\title{Complex chromoelectric polarizability of a heavy-quarkonium resonance: Pole response and channel-resolved effective-theory representation}

\author[bitp]{Arkadiy I. Syamtomov\corref{cor1}}
\ead{arkady.syamtomov@gmail.com}
\cortext[cor1]{Corresponding author}
\affiliation[bitp]{organization={Bogolyubov Institute for Theoretical Physics, National Academy of Sciences of Ukraine},
            city={Kyiv},
            country={Ukraine}}

\begin{abstract}
For a stable compact quarkonium state in the \pNRQCD{} multipole regime,
chromoelectric polarizability arises at leading order from two \(E1\)
transitions through virtual color-octet states.  For an unstable state, we
define the corresponding response through the quadratic displacement of its
complex resonance pole in a weak chromoelectric background, so that its real
and imaginary parts describe field-induced shifts of the pole mass and width.
Starting from a renormalized background-field correlator, we show that the
resonance polarizability is given by the complete two-field kernel normalized
by the energy derivative of the full inverse propagator at the pole.  In an
effective theory with explicitly separated intermediate sectors, the kernel
contains the standard \(E1\)-octet-\(E1\) contribution, the complete source
response of decay channels already coupled at zero field, and local
contributions.  This distinction becomes essential above open-flavor
threshold and reduces to the standard \pNRQCD{} polarizability in the stable,
energy-independent limit.  A simple \(D\bar D\) realization for the
\(\psi(3770)\) gives the complex pole-residue factor
\(1/[1-\Sigma'(z_\psi)]=0.800-0.336\,\ii\), illustrating that threshold
dressing can substantially modify both the magnitude and phase of the
response.  The resulting pole-based relation identifies the quantity required
for a source-dependent coupled-channel determination of the chromoelectric
polarizability of a heavy-quarkonium resonance.
\end{abstract}


\end{frontmatter}

\section{Introduction}
\label{sec:intro}

For a stable state in a weak external field, the static electric
polarizability is defined by the quadratic term in the energy shift,
\begin{equation}
 \Delta E=-\bm d_0\!\cdot\!\bm{\mathcal E}
 -\frac12\alpha_E^{ij}\mathcal E_i\mathcal E_j+\cdots ,
 \label{eq:stable-response-definition}
\end{equation}
where \(\bm d_0\) is the permanent dipole moment,
\(\mathcal E_i\) are the Cartesian components of the external field, and
\(\alpha_E^{ij}\) is the static polarizability tensor; repeated spatial
indices are summed.  For a color-singlet heavy-quarkonium state, the
expectation value of a one-field chromoelectric dipole operator vanishes:
the operator carries an adjoint color index, whereas a singlet state
provides no invariant adjoint vector.  The linear response therefore
vanishes, although the singlet--octet \(E1\) transition remains nonzero.
In the \pNRQCD{} multipole regime the leading singlet response is thus
generated by two chromoelectric dipole transitions,
\[
 S\xrightarrow{E1}O\xrightarrow{E1}S,
\]
through a virtual color-octet sector.  This mechanism originates in the
analyses of Peskin \cite{Peskin1979} and Bhanot--Peskin
\cite{BhanotPeskin1979} and is organized systematically in NRQCD and
\pNRQCD{} \cite{BBL1995,pNRQCD2000,RMP2005}.  For a stable compact state,
the corresponding coefficient is real and reduces to the familiar
chromoelectric polarizability
\cite{Brambilla2016,Voloshin2004,Voloshin2006,Guo2006,Polyakov2018}.

The situation changes once the quarkonium state is coupled to an
open-flavor continuum.  Above threshold the corresponding self-energy
develops an absorptive part, and the unstable state is no longer described
by an isolated normalizable level but by a pole of the analytically
continued propagator,
\begin{equation}
 z_\Phi=M_\Phi-\frac{\ii}{2}\Gamma_\Phi .
 \label{eq:pole-position}
\end{equation}
A weak external field shifts this complex pole and therefore changes both
the resonance mass and width.  The resonance analogue of the usual
polarizability must consequently be defined through the quadratic field
dependence of \(z_\Phi\), rather than through the energy shift of a
normalizable stationary state.

Complex response coefficients for metastable states have a
well-established quantum-mechanical precedent; complex-coordinate methods
have, for example, been used to compute static polarizabilities of
antiprotonic and pionic helium resonances
\cite{KorobovZhong2015,KorobovPionic2015}.  In field theory, complex pole
positions and residues provide the natural characterization of unstable
particles \cite{GegeliaScherer2010}, while resonance matrix elements and
external-field responses can be defined by analytic continuation to the
pole \cite{Bernard2012,RuizdeElvira2017,Lozano2022,Meissner2026}.
Background-field methods for second-order response are likewise well
established for stable states \cite{Bouchard2017,Can2020}, and recent
heavy-hadron chiral-EFT calculations exhibit complex electromagnetic
polarizabilities when open channels are present \cite{DangVector2026}.

The aim here is to construct the quadratic chromoelectric response as a
renormalized QCD background-source pole observable and to determine its
heavy-quarkonium effective-theory representation with the relevant
intermediate sectors kept explicit.  Two difficulties arise.  Near an open
threshold the propagator is energy dependent, so the relation between a
quadratic source insertion and the displacement of the resonance pole is
not the same as for an isolated bound-state level.  Moreover, the external
chromoelectric field can couple the compact quarkonium state to virtual
color-octet configurations, whereas open-flavor channels are already part
of the zero-field dynamics.  A consistent pole response must therefore
specify how these physically different sectors enter and how their source
dependence is converted into a displacement of the complex pole.

Systematic EFT descriptions of quarkonium--open-flavor mixing already
exist in Born--Oppenheimer EFT and related threshold EFTs
\cite{TarrusCastella2022,BrambillaThreshold2026,BrambillaAboveThreshold2026}.
The present construction specifies how a weak chromoelectric source enters
the pole response once such coupled-channel dynamics is supplied.  It is
also complementary to matching analyses in which local chromoelectric and
chromomagnetic quarkonium couplings are reduced to scalar and spin-two
gluonic operators \cite{SyamtomovScalarSpinTwo}, and to the
QCD-to-BOEFT treatment of the source-dependent response of a normalized
stable state \cite{SyamtomovBOEFT}.  Here the response itself is formulated
at an unstable pole.

We first formulate the quadratic response directly at the resonance pole
and derive its normalization for an energy-dependent propagator.  We then
match this pole response onto the heavy-quarkonium effective theory,
separating the virtual color-octet contribution from open-flavor channels
already present in the zero-field dynamics and from local terms.  The
resulting complex response is illustrated for the \(\psi(3770)\) near the
\(D\bar D\) threshold.

\section{Quadratic response of a complex resonance pole}
\label{sec:pole}

\subsection{Pole definition from the source-dependent correlator}

Let \(J_\Phi\) be a renormalized, gauge-invariant current with the quantum
numbers of the quarkonium state of interest, chosen such that its two-point
function contains the corresponding resonance pole with nonzero residue.
We introduce the external source through a smooth one-parameter family of
renormalized background gauge fields \(\bar A_{R,\mu}(\lambda)\).  The
vacuum-normalized two-point function is
\begin{equation}
\begin{aligned}
 G_J(p',p;\bar A_R)=\int \dd^4x\,\dd^4y\,
 e^{\ii p'\cdot x-\ii p\cdot y}
 \frac{\int\mathcal D\varphi\,J_\Phi(x)J_\Phi^\dagger(y)
 e^{\ii S[\varphi;\bar A_R]}}
 {\int\mathcal D\varphi\,e^{\ii S[\varphi;\bar A_R]}} .
\end{aligned}
\label{eq:qcd-correlator}
\end{equation}
The denominator removes source interactions disconnected from the
interpolating currents.  All source derivatives below are taken along the
specified family \(\bar A_{R,\mu}(\lambda)\).  We use the renormalized
background-source variables
\begin{equation}
 \bar{\mathcal A}_\mu^a \equiv g_R \bar A_{R,\mu}^a,
 \qquad
 \mathcal E_i^a \equiv g_R E_{R,i}^a[\bar A_R],
 \label{eq:background-source-variables}
\end{equation}
where \(\bar{\mathcal A}_\mu^a\) is the coupling-weighted background gauge
potential and \(\mathcal E_i^a\) the corresponding chromoelectric component
of the field strength.  The background-field Ward identity fixes the
renormalization of these combinations
\cite{Abbott1981,Abbott1982}.

On the Riemann sheet containing the resonance pole \(z_\Phi\), the
analytically continued rest-frame zero-source correlator has the local form
\begin{equation}
 G_J(z;0)=\frac{\mathcal R_J}{z-z_\Phi}+G_J^{\rm reg}(z),
 \label{eq:pole-factorization}
\end{equation}
where the residue \(\mathcal R_J\) depends on the interpolating current,
whereas the pole position does not.  Consider a weak source for which the
local chromoelectric field takes the form
\(\mathcal E_i^a=\lambda e_i n^a\), with fixed normalized spatial and color
directions \(e_i\) and \(n^a\).  At zero source,
\(\partial z_\Phi/\partial\mathcal E_i^a\) would transform as an adjoint
color vector.  Since no nonzero invariant adjoint vector can be formed from
a color-singlet pole, the linear displacement vanishes.  This does not
forbid the elementary singlet--octet \(E1\) transition.

The leading pole displacement is therefore
\begin{equation}
 z_\Phi(\lambda)
 =
 z_\Phi-\frac{\lambda^2}{2}\alpha_{\Phi,E}(\bm e)
 +O(\lambda^3),
 \label{eq:pole-definition}
\end{equation}
where, for a tensor response,
\(\alpha_{\Phi,E}(\bm e)=
\alpha_{\Phi,E}^{ij}e_i e_j\).
For a stable pole on the physical sheet, \(z_\Phi\) is real and
\cref{eq:pole-definition} reduces to the usual energy-shift definition in
\cref{eq:stable-response-definition}.  For an unstable state with an open
decay channel, the pole generally lies off the real axis,
\[
 z_\Phi=M_\Phi-\frac{\ii}{2}\Gamma_\Phi ,
\]
and the same definition describes the quadratic response of both its mass
and width.

The independence of this coefficient from the interpolating current is
particularly transparent from the pole expansion.  Allowing both the pole
position and its residue to depend on \(\lambda\), and taking the source
derivatives at fixed complex energy \(z\), one finds
\begin{equation}
\begin{aligned}
 \left.\frac{\partial^2G_J}{\partial\lambda^2}\right|_0
 ={}&
 \frac{\mathcal R_J''}{z-z_\Phi}
 +\frac{2\mathcal R_J' z_\Phi'
       +\mathcal R_J z_\Phi''}
       {(z-z_\Phi)^2}
 +\frac{2\mathcal R_J(z_\Phi')^2}
       {(z-z_\Phi)^3}
 +\text{regular}.
\end{aligned}
\label{eq:double-pole-expanded}
\end{equation}
The absence of a linear chromoelectric response for a color-singlet pole
implies \(z_\Phi'(0)=0\), while
\cref{eq:pole-definition} gives
\(z_\Phi''(0)=-\alpha_{\Phi,E}\).  Hence
\begin{equation}
 \left.\frac{\partial^2G_J}{\partial\lambda^2}\right|_0
 =
 \frac{\mathcal R_J''}{z-z_\Phi}
 -\frac{\mathcal R_J\alpha_{\Phi,E}}
       {(z-z_\Phi)^2}
 +\text{regular}.
 \label{eq:double-pole-curvature}
\end{equation}
The source dependence of the residue therefore contributes only to the
simple pole, whereas displacement of the resonance pole produces the
double pole.  For any interpolating current with nonzero residue at the
same pole,
\begin{equation}
 \alpha_{\Phi,E}
 =
 -\frac{\text{coefficient of }(z-z_\Phi)^{-2}}
        {\mathcal R_J}.
 \label{eq:pole-curvature-from-correlator}
\end{equation}
The right-hand side is thus independent of the choice of interpolating
current; invariance under an overall rescaling of that current is manifest,
since the numerator and denominator acquire the same factor.

\subsection{Pole perturbation theory and residue normalization}

Let \(z_\Phi(\lambda)\) denote an isolated resonance pole.  Equivalently,
the renormalized inverse propagator has a zero mode,
\(
\mathcal D_\Phi^{-1}(z_\Phi(\lambda);\lambda)
\ket{R_\Phi(\lambda)}=0
\).
At zero source we introduce right and left pole vectors satisfying
\(
\mathcal D_{\Phi,0}^{-1}(z_\Phi)\ket{R_\Phi}=0
\)
and
\(
\bra{L_\Phi}\mathcal D_{\Phi,0}^{-1}(z_\Phi)=0
\),
with normalization \(\langle L_\Phi|R_\Phi\rangle=1\).
Because the inverse propagator analytically continued to the resonance
sheet is generally non-Hermitian, the two vectors form a biorthogonal pair
and need not be related by Hermitian conjugation.  Analytic perturbation
theory for an isolated simple characteristic value is standard
\cite{Kato1995}.

Near the pole the propagator has the singular part
\begin{equation}
 \mathcal D_\Phi(z)
 \underset{z\to z_\Phi}{=}
 \frac{\ket{R_\Phi}\bra{L_\Phi}}
 {(z-z_\Phi)\,\mathcal N_\Phi}
 +\text{regular},
 \qquad
 \mathcal N_\Phi\equiv
 \bra{L_\Phi}
 \partial_z\mathcal D_{\Phi,0}^{-1}(z_\Phi)
 \ket{R_\Phi}.
 \label{eq:propagator-residue}
\end{equation}
The factor \(\mathcal N_\Phi^{-1}\) is therefore the normalization of the
pole residue associated with the energy dependence of the inverse
propagator.

All source derivatives below are taken at fixed complex energy \(z\), or
at fixed residual energy \(\varepsilon\) after the heavy-quark mass has
been separated.  We write
\begin{equation}
 \mathcal D_\Phi^{-1}(z;\lambda)
 =
 \mathcal D_{\Phi,0}^{-1}(z)
 +\lambda\,\mathcal V_{\Phi,E}(z)
 +\frac{\lambda^2}{2}\mathcal V_{\Phi,EE}(z)
 +O(\lambda^3),
 \label{eq:inverse-expansion}
\end{equation}
where
\(
\mathcal V_{\Phi,E}
 =\partial_\lambda\mathcal D_\Phi^{-1}|_{\lambda=0}
\)
and
\(
\mathcal V_{\Phi,EE}
 =\partial_\lambda^2\mathcal D_\Phi^{-1}|_{\lambda=0}
\).
The motion of the pole itself is not included in these partial
derivatives.

Differentiating the pole equation once and projecting with
\(\bra{L_\Phi}\) gives
\(
z_\Phi'(0)\mathcal N_\Phi+
\bra{L_\Phi}\mathcal V_{\Phi,E}\ket{R_\Phi}=0
\).
For the chromoelectric response of a color-singlet pole,
\(
\bra{L_\Phi}\mathcal V_{\Phi,E}\ket{R_\Phi}=0
\),
because a one-field insertion carries an adjoint color index and cannot
produce an invariant adjoint vector between singlet pole states.
Consequently \(z_\Phi'(0)=0\), in agreement with the quadratic definition
in \cref{eq:pole-definition}.  This condition concerns only the
pole-projected matrix element: the one-field kernel itself may still
connect the pole to other configurations.

Those transitions contribute at second order.  After one source insertion
the pole state may propagate through a complementary intermediate sector
before a second insertion returns it to the pole subspace.  The inverse
propagator cannot be inverted on the full space at \(z=z_\Phi\), since it
has the pole vector as a zero mode.  We therefore separate the pole direction and define
\begin{equation}
\begin{aligned}
 P_\Phi&=\ket{R_\Phi}\bra{L_\Phi},
 \qquad
 Q_\Phi=1-P_\Phi,\\
 S_\Phi(z)&=
 Q_\Phi
 \bigl[
 Q_\Phi\mathcal D_{\Phi,0}^{-1}(z)Q_\Phi
 \bigr]^{-1}
 Q_\Phi,
 \qquad
 S_\Phi\equiv S_\Phi(z_\Phi).
\end{aligned}
\label{eq:reduced-propagator}
\end{equation}
For an isolated simple pole, the inverse in the \(Q_\Phi\) subspace is
regular at \(z=z_\Phi\).  The reduced resolvent \(S_\Phi(z)\) describes
propagation in the complementary subspace with the resonance pole itself
removed; its value \(S_\Phi(z_\Phi)\) enters the static second-order
response.

Choosing the first-order correction to the pole vector such that
\(\bra{L_\Phi}R_\Phi^{(1)}\rangle=0\), the projected first-order equation
gives
\(
\ket{R_\Phi^{(1)}}=
-S_\Phi\mathcal V_{\Phi,E}\ket{R_\Phi}
\).
Substitution into the second derivative of the pole equation then yields
the general quadratic response
\begin{equation}
 \boxed{
 \alpha_{\Phi,E}=
 \frac{\bra{L_\Phi}
 \left[
 \mathcal V_{\Phi,EE}
 -2\mathcal V_{\Phi,E}S_\Phi\mathcal V_{\Phi,E}
 \right]
 \ket{R_\Phi}}
 {\bra{L_\Phi}
 \partial_z\mathcal D_{\Phi,0}^{-1}(z_\Phi)
 \ket{R_\Phi}} .}
 \label{eq:general-pole-curvature}
\end{equation}
The first term is the direct two-field response.  The second describes two
successive one-field transitions through intermediate configurations and
is the usual reduced-resolvent contribution of second-order perturbation
theory.

If those complementary sectors have already been eliminated, their
iterated contribution is contained in the effective quadratic kernel of
the reduced theory.  Denoting this complete reduced kernel by
\(\mathcal V_{\Phi,EE}^{\rm red}\), the same result takes the simpler form
\begin{equation}
 \boxed{
 \alpha_{\Phi,E}=
 \frac{\bra{L_\Phi}
 \mathcal V_{\Phi,EE}^{\rm red}(z_\Phi)
 \ket{R_\Phi}}
 {\bra{L_\Phi}
 \partial_z\mathcal D_{\Phi,0}^{-1}(z_\Phi)
 \ket{R_\Phi}} .}
 \label{eq:pole-curvature}
\end{equation}
Thus the reduced form does not follow merely from the absence of a linear
pole displacement; it applies when the intermediate one-field response
has been incorporated consistently into the quadratic kernel.  For an
energy-independent bound-state Hamiltonian,
\(\mathcal D^{-1}=z-H\), the denominator reduces to unity.

For a degenerate but semisimple pole, the residue normalization and the
quadratic response become matrices in the pole subspace.  With
biorthogonal vectors
\(\{\ket{R_\lambda},\bra{L_\lambda}\}\), define
\begin{equation}
 (Z_\Phi^{-1})_{\lambda'\lambda}
 =
 \bra{L_{\lambda'}}
 \partial_z\mathcal D_{\Phi,0}^{-1}(z_\Phi)
 \ket{R_\lambda},
 \qquad
 (K_{EE}(\bm e))_{\lambda'\lambda}
 =
 \bra{L_{\lambda'}}
 \mathcal K_{\Phi,EE}(\bm e)
 \ket{R_\lambda},
 \label{eq:multiplet-matrices}
\end{equation}
where \(\mathcal K_{\Phi,EE}\) denotes the complete quadratic kernel,
equal in the explicit-channel representation to
\(
\mathcal V_{\Phi,EE}
-2\mathcal V_{\Phi,E}S_\Phi\mathcal V_{\Phi,E}
\).
Since the external field can mix the degenerate pole vectors, the
quadratic shifts are obtained from the generalized eigenvalue problem
\(
[K_{EE}(\bm e)-\alpha_E(\bm e)Z_\Phi^{-1}]\,c=0
\),
or equivalently
\begin{equation}
 \det\!\left[
 K_{EE}(\bm e)-\alpha_E(\bm e)Z_\Phi^{-1}
 \right]=0 .
 \label{eq:multiplet-response}
\end{equation}
For a spin-one resonance such as the \(\psi(3770)\), the general
formulation accommodates polarization-dependent quadratic shifts.  In the
application below we retain only the scalar, spin-averaged component.

The role of the residue normalization is particularly transparent for a
single scalar degree of freedom.  If
\(
D^{-1}(z;\lambda)
=z-E_0-\Sigma(z)
+\lambda^2\mathcal V_{EE}^{\rm red}(z)/2+\cdots
\),
then the zero-source pole obeys
\(z_\Phi-E_0-\Sigma(z_\Phi)=0\), and
\cref{eq:pole-curvature} gives
\begin{equation}
 \boxed{
 \alpha_{\Phi,E}
 =
 \frac{\mathcal V_{EE}^{\rm red}(z_\Phi)}
      {1-\Sigma'(z_\Phi)} .}
 \label{eq:minimal-alpha}
\end{equation}
The factor \(1/[1-\Sigma'(z_\Phi)]\) is the scalar form of the pole-residue
normalization and accounts for the energy dependence generated by the
open-channel dressing.

\subsection{Mass and width shifts}

For a real static field, global color invariance reduces the quadratic color structure to \(\delta^{ab}\alpha_{\Phi,E}^{ij}\), so
\begin{equation}
 z_\Phi(\cE)=z_\Phi-\frac12\alpha_{\Phi,E}^{ij}\mathcal E_i^a\mathcal E_j^a+O(\cE^3).
 \label{eq:static-definition}
\end{equation}
Since
\(
z_\Phi=M_\Phi-\ii\Gamma_\Phi/2
\),
comparison of the real and imaginary parts of the pole shift gives
\begin{align}
 \Delta M_\Phi
 &=
 -\frac12
 \operatorname{Re}\alpha_{\Phi,E}^{ij}
 \mathcal E_i^a\mathcal E_j^a,
 &
 \Delta\Gamma_\Phi
 &=
 \operatorname{Im}\alpha_{\Phi,E}^{ij}
 \mathcal E_i^a\mathcal E_j^a .
 \label{eq:mass-width}
\end{align}
The real and imaginary parts of the complex polarizability therefore govern,
respectively, the shifts of the pole mass and width.

\section{Static and frequency-dependent response}
\label{sec:dynamic}

The pole definition of the previous section does not depend on the
particular realization of the weak source.  To connect it with the
chromoelectric polarizability used in the heavy-quark multipole expansion,
we now specify the static and long-wavelength limits of the background
field.

For the static response, the chromoelectric background is taken to vary
slowly across the compact heavy-quark core.  If \(a_\Phi\) denotes the
characteristic size of that core and \(R\sim |\bm q|^{-1}\) the spatial
variation scale of the background, the multipole expansion requires
\[
 a_\Phi\ll R .
\]
Since the polarizability is defined by derivatives at vanishing source,
the construction does not require an exactly uniform non-Abelian field
over all space.  It is sufficient that the background be smooth over the
quarkonium system, with the resonance pole remaining isolated as the
long-wavelength limit is taken.

Zero four-momentum can be approached in two physically distinct ways.
The static response corresponds to a time-independent field,
\(q^\mu=(0,\bm q)\), followed by \(|\bm q|\to0\).  The zero-frequency
limit of a spatially homogeneous time-dependent field instead corresponds
to \(q^\mu=(\omega,\bm 0)\) with \(\omega\to0\).  The two limits coincide
when the analytically continued two-field amplitude is regular in a
neighborhood of \(q^\mu=0\).  Nearby continuum thresholds can introduce
nonanalytic energy dependence, in which case the order of the limits may
matter.  Throughout this work, the static polarizability refers to the
first limit.

For a monochromatic background, write
\begin{equation}
 \mathcal E_i^a(t)
 =
 n^a\left[
 \mathcal E_{+,i}e^{-\ii\omega t}
 +\mathcal E_{-,i}e^{+\ii\omega t}
 \right],
 \qquad
 \mathcal E_{-,i}=\mathcal E_{+,i}^{*}
 \quad\text{for a real field}.
 \label{eq:periodic-field}
\end{equation}
We follow the Floquet quasienergy branch that approaches the unperturbed
resonance pole when the field is switched off,
\(
z_\Phi^{\rm q}\to z_\Phi
\)
as \(\mathcal E_\pm\to0\).  Its leading quadratic displacement defines
the frequency-dependent response,
\begin{equation}
 z_\Phi^{\rm q}(\mathcal E_-,\mathcal E_+)
 =
 z_\Phi
 -\alpha_{\Phi,E}^{ij}(\omega)
  \mathcal E_{-,i}\mathcal E_{+,j}
 +O(\mathcal E^3).
 \label{eq:dynamic-definition}
\end{equation}

When one-field transitions to complementary sectors are retained
explicitly, absorption and emission of the external frequency lead to
intermediate propagation at the shifted complex energies
\(z+\omega\) and \(z-\omega\).  The complete quadratic kernel is then
\begin{equation}
\begin{aligned}
 \mathcal K_{\Phi,EE}^{ij}(z,\omega)
 ={}&
 \mathcal V_{\Phi,EE}^{ij}(z,\omega)
 -\mathcal V_{\Phi,E,-}^{i}
  S_\Phi(z+\omega)
  \mathcal V_{\Phi,E,+}^{j}
\\
 &-\mathcal V_{\Phi,E,+}^{j}
  S_\Phi(z-\omega)
  \mathcal V_{\Phi,E,-}^{i}.
\end{aligned}
\label{eq:dynamic-general-kernel}
\end{equation}
At \(\omega=0\), for a single static source direction, this reduces to
the second-order kernel in
\cref{eq:general-pole-curvature}.  If the complementary sectors have
instead already been eliminated, their iterated contribution is contained
in the reduced two-field kernel, and one simply has
\[
 \mathcal K_{\Phi,EE}^{ij}
 =
 \mathcal V_{\Phi,EE}^{\rm red,\,ij}.
\]
In particular, this reduced form does not imply the absence of the
underlying singlet--octet \(E1\) transition; its effect has been absorbed
into the quadratic singlet kernel.

The mixed derivative with respect to the positive- and negative-frequency
source components is the pole part of the corresponding two-current
response, including local contact terms.  Interchanging the two source
derivatives gives
\[
 \alpha_{\Phi,E}^{ij}(\omega)
 =
 \alpha_{\Phi,E}^{ji}(-\omega)
\]
when both amplitudes are continued on the same Riemann sheet.  For an
isotropic scalar response, \(\alpha_{\Phi,E}(\omega)\) is therefore an
even function of \(\omega\).

The perturbative form above assumes that the intermediate resolvents
remain regular at \(z_\Phi\pm\omega\).  If one of these energies approaches
another isolated pole, the corresponding states can no longer be treated
as a nondegenerate virtual intermediate sector; the pole subspace must be
enlarged and the source-induced mixing diagonalized.  If instead
\(z_\Phi\pm\omega\) approaches a continuum threshold, the reduced
resolvent develops the associated branch-point nonanalyticity and the
relevant continuum dynamics must be retained explicitly rather than
represented by a regular perturbative denominator.

\section{Channel reduction and effective-theory representation}
\label{sec:matching}

\subsection{Octet response and effective-theory degrees of freedom}

The preceding sections defined the quadratic pole response without
specifying the microscopic origin of the intermediate states.  We now
identify those states for heavy quarkonium in the pNRQCD multipole
expansion.  When the hierarchy
\[
 m_Q \gg m_Q v \gg m_Q v^2
\]
is well separated, the compact \(Q\bar Q\) system admits a description
in terms of singlet and color-octet fields interacting with ultrasoft
gluons.  In the weak-coupling regime, where the soft scale \(m_Qv\) is
perturbative, the leading chromoelectric coupling is \cite{pNRQCD2000,RMP2005}
\begin{equation}
 \mathcal L_{E1}=V_A\!\int\!\dd^3\bm r\,
 \mathrm{Tr}\!\left[
 O^\dagger\bm r\!\cdot\!\cE\,S
 +S^\dagger\bm r\!\cdot\!\cE\,O\right].
 \label{eq:E1-lagrangian}
\end{equation}
Writing the matrix-valued pNRQCD fields as
\[
 S=\frac{\mathbf 1_c}{\sqrt{N_c}}\,S_s,
 \qquad
 O=\frac{T^a}{\sqrt{T_F}}\,O^a,
 \qquad
 \mathrm{Tr}(T^aT^b)=T_F\delta^{ab},
\]
where \(S_s\) is the color-singlet component and \(O^a\) the adjoint
color-octet components, the two time orderings give the nonlocal dipole
kernel
\begin{equation}
 \mathcal V_{o,EE}^{ij}(\varepsilon,\omega)=
 \frac{T_F}{N_c}V_A^2\left[
 r^i\frac{1}{h_o-\varepsilon-\omega}r^j
 +r^j\frac{1}{h_o-\varepsilon+\omega}r^i
 \right].
 \label{eq:dynamic-pnrqcd}
\end{equation}
The explicit color-octet realization is specific to weakly coupled
\pNRQCD{}.  More generally, the complementary sector entering the
second-order pole response is whatever set of gluonic or hybrid
excitations remains dynamical at the scale of interest.  The pole-response
formula derived above is unchanged; only the effective-theory
representation of the intermediate sector differs.

\subsection{Open channels and the Feshbach--Schur reduction}

An open-flavor decay channel differs qualitatively from the virtual octet
because it can mix with the compact state already in the absence of the
external source.  To describe this mixing, split the channel space into
the retained compact-singlet sector \(S\) and the complementary
open-channel sector \(X\).  The block \(\Gamma_{XX}\) is the inverse
two-point function restricted to the \(X\) sector, and we denote its
inverse by
\[
 G_X\equiv\Gamma_{XX}^{-1}.
\]
Thus \(G_X\) describes propagation within the complementary sector
between transitions from and back to the compact state.

In residual-energy variables
\(\varepsilon=z-2m_Q^{\mathcal S_m}(\mu_m)\), the full inverse
two-point function has the block form
\begin{equation}
 \bm\Gamma(\varepsilon;\cE)=
 \begin{pmatrix}
  \Gamma_{SS} & \Gamma_{SX}\\
  \Gamma_{XS} & \Gamma_{XX}
 \end{pmatrix}.
 \label{eq:coupled-inverse}
\end{equation}
For a time-dependent background these blocks are understood as energy-space kernels, and their products include the corresponding energy convolution; a monochromatic source connects residual energies differing by \(\omega\).
The standard Feshbach--Schur reduction \cite{Feshbach1958,Feshbach1962} gives the exact inverse propagator in the retained sector,
\begin{equation}
 \mathcal D_\Phi^{-1}=\Gamma_{SS}-\Gamma_{SX}G_X\Gamma_{XS}.
 \label{eq:effective-singlet-inverse}
\end{equation}
The second term is the self-energy induced by excursions into the
complementary sector,
\[
 \Sigma_X=\Gamma_{SX}G_X\Gamma_{XS},
\]
so that
\(
\mathcal D_\Phi^{-1}=\Gamma_{SS}-\Sigma_X
\).
Systematic realizations of such a coupled quarkonium--open-flavor sector have been constructed in Born--Oppenheimer EFT and threshold EFT, where quarkonium fields mix with heavy-meson or tetraquark configurations and generate energy-dependent mass shifts and widths \cite{TarrusCastella2022,BrambillaThreshold2026,BrambillaAboveThreshold2026}.  Equation~\eqref{eq:effective-singlet-inverse} is used here at the operator level: once a particular coupled-channel EFT and its source couplings are specified, its blocks supply the corresponding \(\Gamma\)'s.

This identity also exposes the physical distinction from the virtual octet.  If \(\Gamma_{SX}(0)\) and \(\Gamma_{XS}(0)\) are already nonzero, the quadratic source response receives contributions from source insertions on the channel propagator, on either mixing vertex, and from genuine two-source contacts.  Defining
\begin{equation}
 \Sigma_{\rm op}(\varepsilon;\cE)
 \equiv [\Gamma_{SX}G_X\Gamma_{XS}]_{\rm open},
\end{equation}
the source derivatives below are partial derivatives at fixed residual energy \(\varepsilon\), in accordance with the convention stated in \cref{sec:pole}.  The whole set is then summarized without ambiguity by
\begin{equation}
 \boxed{
 \mathcal V_{{\rm op},EE}^{ij}(\varepsilon,\omega)=
 -\left.
 \frac{\partial^2\Sigma_{\rm op}}
 {\partial\mathcal E_{-,i}\partial\mathcal E_{+,j}}
 \right|_{\cE=0}.}
 \label{eq:open-channel}
\end{equation}
The product rule for the mixed derivative \(\partial_{-,i}\partial_{+,j}(\Gamma_{SX}G_X\Gamma_{XS})\) then generates all of these insertions automatically, including terms proportional to the undifferentiated zero-field mixing vertices.  By contrast, the leading singlet--octet term starts from vanishing zero-field singlet--octet mixing, so both transitions are created by the external field.  This is the specific reason why the familiar \(E1\,G_o\,E1\) reduction applies to the octet but not to a generic open decay channel.

\subsection{Local terms and the channel-resolved pole response}

Short-distance modes and high-gap excitations already integrated out of the explicit channel space generate a local two-field vertex \(\mathcal V_{{\rm loc},EE}^{ij}\).  Let \(\Delta_n\equiv E_n-z_\Phi\) denote the gap of an eliminated excitation.  When \(|\omega|\ll|\Delta_n|\), the sum of the two time orderings is analytic,
\begin{equation}
 \frac{1}{z_\Phi+\omega-E_n}
 +\frac{1}{z_\Phi-\omega-E_n}
 =-\frac{2}{\Delta_n}\left[1+\frac{\omega^2}{\Delta_n^2}
 +O\!\left(\frac{\omega^4}{\Delta_n^4}\right)\right].
 \label{eq:high-gap-expansion}
\end{equation}
The odd powers cancel in the crossing-even scalar response, and the remaining analytic series is represented by local low-energy operators.

In the color-covariant singlet reduction used here, where the complete reduced one-field kernel vanishes, \(\mathcal K_{\Phi,EE}^{ij}=\mathcal V_{\Phi,EE}^{ij}\).  When the explicitly retained octet/hybrid and open-channel sectors form a nonoverlapping partition of the resolved complementary space, this two-field kernel may be organized as
\begin{equation}
 \mathcal V_{\Phi,EE}^{ij}
 =\mathcal V_{{\rm dip},EE}^{ij}
 +\mathcal V_{{\rm op},EE}^{ij}
 +\mathcal V_{{\rm loc},EE}^{ij} .
 \label{eq:decomposition}
\end{equation}
If the explicit sectors mix, their combined contribution is defined directly by differentiating \cref{eq:effective-singlet-inverse}; the decomposition in \cref{eq:decomposition} is then only a bookkeeping choice.

Combining this kernel with the pole normalization gives the channel-resolved pole-response relation
\begin{equation}
 \boxed{
 \alpha_{\Phi,E}^{ij}(\omega)=
 \frac{
 \bra{L_\Phi}
 [\mathcal K_{\Phi,EE}^{ij}(\varepsilon,\omega;\mu_f,\mu_m)]_{\varepsilon_\Phi}
 \ket{R_\Phi}}
 {\bra{L_\Phi}
 [\partial_\varepsilon\mathcal D_\Phi^{-1}(\varepsilon;0;\mu_f,\mu_m)]_{\varepsilon_\Phi}
 \ket{R_\Phi}},
 \quad
 \varepsilon_\Phi=z_\Phi-2m_Q^{\mathcal S_m}(\mu_m).}
 \label{eq:main-result}
\end{equation}
Here \(\mu_f\) denotes the EFT factorization scale separating the explicitly retained low-energy sectors from local operators, while \(m_Q^{\mathcal S_m}(\mu_m)\) is the heavy-quark mass in a specified short-distance scheme \(\mathcal S_m\) at scale \(\mu_m\).  The generalized kernel \(\mathcal K_{\Phi,EE}^{ij}\) is given by \cref{eq:dynamic-general-kernel}, with the same definition in residual-energy variables; in the color-covariant singlet reduction used for the decomposition in \cref{eq:decomposition}, it reduces to \(\mathcal V_{\Phi,EE}^{ij}\).  Equation~\eqref{eq:main-result} is an operator-level effective-theory representation for a chosen nonoverlapping channel partition; a concrete open-flavor EFT must still supply the corresponding channel fields, couplings, power counting, and source operators.

\subsection{Stable-state limit}

A useful consistency check is obtained by removing the open channels, taking an energy-independent singlet Hamiltonian, and neglecting higher-order local two-field operators.  For a nondegenerate \(S\)-wave state \(\ket{\phi}\) with \(h_s\ket{\phi}=E_\phi\ket{\phi}\), \cref{eq:dynamic-pnrqcd,eq:main-result} at \(\omega=0\) gives
\begin{equation}
 \boxed{
 \alpha_{\phi,E}(0)=
 \frac{2T_F}{3N_c}V_A^2
 \bra{\phi}\bm r\!\cdot\!\frac{1}{h_o-E_\phi}\bm r\ket{\phi}.}
 \label{eq:stable-polarizability}
\end{equation}
This is the standard \pNRQCD{} chromoelectric polarizability \cite{Brambilla2016}.  Thus the resonance formula retains the familiar stable limit while adding exactly the structures required by an energy-dependent pole and by channels already open at zero field.

\section{A \texorpdfstring{\(\psi(3770)\)}{psi(3770)} threshold realization}
\label{sec:numerical}

The \(\psi(3770)\) provides a useful benchmark because its pole lies close to the \(D\bar D\) thresholds.  The coupled-channel analysis of Ref.~\cite{Hanhart2024} gives
\(M_\psi=3778.8(3)\,\mathrm{MeV}\) and
\(\Gamma_\psi=25.0(5)\,\mathrm{MeV}\), with the width dominated by nearby \(D\bar D\) channels.  The scalar model below represents the spin-averaged (scalar) component of the multiplet response in \cref{eq:multiplet-response}.  This benchmark tests the pole-residue normalization formula and open-threshold kinematics; it does not assume that the physical \(\psi(3770)\) is quantitatively described by weak-coupling \pNRQCD{}.  An absolute chromoelectric polarizability requires the appropriate source-dependent coupled-channel dynamics.

\subsection{One-threshold benchmark}

Approximate the charged and neutral channels initially by a single effective \(P\)-wave threshold \(E_{\rm th}=3.735\,\mathrm{GeV}\),
\begin{align}
 D_{\rm II}^{-1}(z,\lambda)
 &=z-E_0-\Sigma_{\rm II}(z,\lambda)
 +\frac{\lambda^2}{2}\mathcal V_{\rm core},
 \nonumber\\
 \Sigma_{\rm II}(z,\lambda)
 &=-\ii\gamma[z-E_{\rm th}(\lambda)]_{\rm II}^{3/2}.
 \label{eq:numerical-model}
\end{align}
For \(w=z-E_{\rm th}(\lambda)\) we use
\begin{equation}
 [w]_{\rm II}^{3/2}=|w|^{3/2}
 \exp\!\left(\frac{3\ii}{2}\theta_{\rm II}\right),
 \qquad -2\pi<\theta_{\rm II}=\arg_{\rm II}w<0 .
 \label{eq:second-sheet-branch}
\end{equation}
Requiring the pole \(z_\psi=3.7788-0.0125\,\ii\,\mathrm{GeV}\) while keeping \(E_0\) real fixes
\begin{equation}
 \gamma=1.4064\,\mathrm{GeV}^{-1/2},
 \qquad E_0=3.78434\,\mathrm{GeV}.
 \label{eq:numerical-fit}
\end{equation}
At the pole,
\begin{align}
 \Sigma_{\rm II}'(z_\psi)&=-0.06238-0.44588\,\ii,
 \nonumber\\
 \mathcal Z_\psi\equiv\frac{1}{1-\Sigma_{\rm II}'(z_\psi)}
 &=0.80031-0.33589\,\ii
 =0.86794e^{-0.3974\ii}.
 \label{eq:numerical-residue}
\end{align}
For a real positive field-independent core vertex this factor reduces the real component of the response by \(19.97\%\), reduces its modulus by \(13.21\%\), and generates a phase of \(-22.8^\circ\).

The independent source response of the open channel can be displayed by writing
\(E_{\rm th}(\lambda)=E_{\rm th}-\beta_{\rm op}\lambda^2/2\) and
\(b=\beta_{\rm op}/\mathcal V_{\rm core}\).  Then
\begin{equation}
 \frac{\alpha_{\psi,E}}{\mathcal V_{\rm core}}
 =\frac{1-b\Sigma_{\rm II}'(z_\psi)}{1-\Sigma_{\rm II}'(z_\psi)} .
 \label{eq:numerical-alpha}
\end{equation}
At \(b=0\), \cref{eq:numerical-alpha} reduces to the residue factor in \cref{eq:numerical-residue}.  A nonzero \(b\) changes both its magnitude and phase through the source response of the already-open channel.  For example, the purely illustrative choice \(b=0.5\) gives \(\alpha_{\psi,E}/\mathcal V_{\rm core}=0.9002-0.1679\,\ii\); this value is used below only as an algebraic test of the open-channel term, not as a phenomenological input.

Under an energy-dependent redefinition of the effective resonance field, \(\mathcal Z_\psi\) and the corresponding two-field vertex transform together.  Their consistently transformed product, equivalently the complete ratio in \cref{eq:main-result}, is the pole property.

\subsection{Two physical thresholds}

The physical neutral and charged thresholds provide a direct test of the one-threshold simplification.  Using the 2025 Particle Data Group values \cite{PDG2025},
\begin{equation}
 E_n=2m_{D^0}=3.72968\,\mathrm{GeV},\qquad
 E_c=2m_{D^+}=3.739324\,\mathrm{GeV}.
\end{equation}
The charged and neutral couplings obtained in Ref.~\cite{Hanhart2024} are compatible with isospin symmetry, so an equal-coupling two-threshold extension is
\begin{equation}
 \Sigma_{\rm II}^{(2)}(z)=-\ii\gamma_2\left(
 [z-E_n]_{\rm II}^{3/2}+[z-E_c]_{\rm II}^{3/2}\right).
 \label{eq:two-threshold-self-energy}
\end{equation}
Refitting \(\gamma_2\) and a real bare energy to the same pole yields
\begin{align}
 \gamma_2&=0.68782\,\mathrm{GeV}^{-1/2},
 &E_0^{(2)}&=3.78424\,\mathrm{GeV},
 \nonumber\\
 \Sigma_{\rm II}^{(2)\prime}(z_\psi)&=-0.06093-0.43796\,\ii,
 &\mathcal Z_\psi^{(2)}&=0.80533-0.33245\,\ii .
 \label{eq:two-threshold-result}
\end{align}
Relative to the original one-threshold benchmark at \(E_{\rm th}=3.735\,\mathrm{GeV}\), the physical two-threshold model changes the complex residue factor by \(0.70\%\).  That comparison also contains a small shift of the threshold centroid.  To isolate the splitting itself, define the coincident-threshold control
\begin{equation}
 \bar E_{\rm th}=\frac{E_n+E_c}{2}=3.734502\,\mathrm{GeV}.
 \label{eq:average-threshold}
\end{equation}
Refitting the one-threshold ansatz at \(\bar E_{\rm th}\) to the same pole gives
\begin{align}
 \gamma_{\rm av}&=1.38177\,\mathrm{GeV}^{-1/2},
 &E_{0}^{\rm av}&=3.78427\,\mathrm{GeV},
 \nonumber\\
 \Sigma_{\rm av}'(z_\psi)&=-0.06096-0.44047\,\ii,
 &\mathcal Z_\psi^{\rm av}&=0.80397-0.33378\,\ii .
 \label{eq:average-threshold-result}
\end{align}
To quantify the splitting effect, define
\begin{equation}
 \delta_Z\equiv
 \frac{|\mathcal Z_\psi^{(2)}-\mathcal Z_\psi^{\rm av}|}
 {|\mathcal Z_\psi^{\rm av}|}.
 \label{eq:delta-z}
\end{equation}
Using \cref{eq:average-threshold-result,eq:two-threshold-result} gives \(\delta_Z=2.2\times10^{-3}\), or \(0.22\%\).  Thus, within this equal-coupling \(P\)-wave ansatz, the physical neutral--charged threshold splitting itself is a sub-percent effect.  The absolute value of \(\Sigma'(z_\psi)\), however, remains sensitive to the broader coupled-channel parametrization, including analytic self-energy terms, form factors, and nearby-state mixing.

\subsection{Direct curvature and continuity checks}

As a numerical check of the source expansion, we solve the full source-dependent pole equation with \(\mathcal V_{\rm core}=1\,\mathrm{GeV}^{-3}\) and form
\begin{equation}
 \alpha_{\rm FD}(\lambda)=-\frac{2[z_\psi(\lambda)-z_\psi(0)]}{\lambda^2}.
 \label{eq:fd-definition}
\end{equation}
For \(b=0\), the relative difference from the analytic result decreases from \(3.73\times10^{-4}\) at \(\lambda=0.020\,\mathrm{GeV}^2\) to \(5.82\times10^{-6}\) at \(0.0025\,\mathrm{GeV}^2\).  As an algebraic test of the open-channel term, the illustrative choice \(b=0.5\) gives a decrease from \(8.83\times10^{-5}\) to \(1.38\times10^{-6}\) over the same range.  Each halving of \(\lambda\) reduces the error by approximately a factor of four, confirming the \(O(\lambda^4)\) remainder of this explicitly even-in-\(\lambda\) benchmark model.

\begin{figure}[ht]
  \centering
  \includegraphics[width=0.78\textwidth]{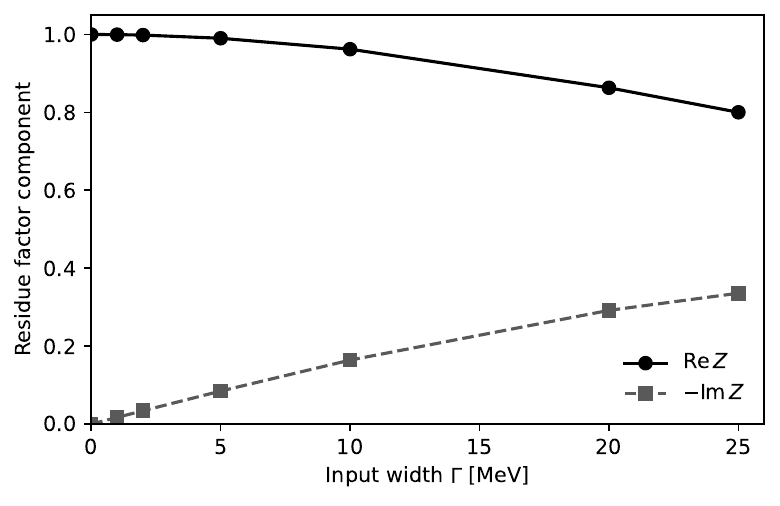}
  \caption{Residue factor in the calibrated one-threshold model as the input width is reduced while the pole mass and effective threshold are kept fixed and \(\gamma\) and \(E_0\) are refitted at each point.  The plotted imaginary component is \(-\operatorname{Im}\mathcal Z\).  As \(\Gamma\to0\), the fitted open-channel coupling vanishes and \(\operatorname{Re}\mathcal Z\to1\), \(\operatorname{Im}\mathcal Z\to0\), showing the continuous zero-width limit of this ansatz.}
  \label{fig:width-scan}
\end{figure}

A second continuity check keeps the pole mass and effective threshold fixed, decreases the input width, and refits \(\gamma\) and \(E_0\) at each point.  The scan in \cref{fig:width-scan} displays the continuous zero-width limit of the calibrated one-threshold ansatz as the fitted open-channel coupling is removed.

The benchmark therefore determines the size and phase of the residue normalization for the specified threshold dynamics.  A quantitative resonance polarizability requires a source-dependent coupled-channel calculation of the open-flavor sector together with the compact-state two-field vertex and local matching terms.  The BOEFT constructions of Refs.~\cite{TarrusCastella2022,BrambillaThreshold2026,BrambillaAboveThreshold2026} provide examples of systematic channel dynamics into which such source couplings can be embedded.

\section{Renormalization, analytic continuation, and domain of applicability}
\label{sec:renorm}

The source variable \(\mathcal E_i^a=g_RE_{R,i}^a\) is the chromoelectric curvature of the renormalized background connection \(\mathcal A_\mu^a=g_R\bar A_{R,\mu}^a\).  In the background-field scheme the Ward identity
\begin{equation}
 Z_g Z_{\bar A}^{1/2}=1
\end{equation}
fixes this source normalization \cite{Abbott1981,Abbott1982}.  Local source contacts attached only to an interpolating current modify the current residue and hence the simple-pole part of a differentiated correlator.  Bulk two-field contacts modify the inverse propagator itself and contribute to \(\mathcal V_{{\rm loc},EE}\); they therefore survive the double-pole extraction in \cref{eq:double-pole-curvature}.

Potentials, self-energies, local vertices, and the residual energy separately depend on the EFT factorization scale and on the heavy-quark mass convention.  The physical pole energy \(z_\Phi\) is fixed, while a change of mass scheme shifts \(\varepsilon_\Phi=z_\Phi-2m_Q^{\mathcal S_m}(\mu_m)\) together with the EFT kernels.  The ratio in \cref{eq:main-result} identifies the combination in which a concrete matched calculation must verify cancellation of these dependences, up to higher-order truncation terms.

The pole formula assumes that the selected resonance remains an isolated simple pole under sufficiently weak sources and that the correlator and its source derivatives admit continuation to the same local second-sheet neighborhood.  This is the standard local analytic setting in which pole observables are defined.  If a source-dependent threshold crosses the continuation path or if the pole becomes degenerate with another nearby state, the corresponding coupled channels must be treated together.

For Euclidean finite-volume calculations, a resonance pole is not a single transfer-matrix eigenvalue.  The relevant input is instead the source dependence of the finite-volume spectrum or of the corresponding two-current amplitude, followed by the usual finite-volume reconstruction and continuation to the pole.  Existing first-order external-field resonance relations \cite{RuizdeElvira2017,Lozano2022,Meissner2026} provide the linear analogue; the quadratic relation derived here identifies the two-field quantity and its pole normalization.

The \pNRQCD{} interpretation additionally requires a compact heavy-quark core and a separation between its internal momentum and binding-energy scales.  If the resonance is dominated by an extended hadronic configuration, the same pole definition remains meaningful, but the particular singlet--octet multipole realization in \cref{eq:dynamic-pnrqcd} need not be the appropriate low-energy description.

\section{Conclusions}
\label{sec:conclusions}

In conclusion, we have shown that the chromoelectric polarizability of an
unstable heavy-quarkonium state is naturally defined through the quadratic
displacement of its complex resonance pole.  The corresponding response is
not given by the two-field interaction alone: it is normalized by the energy
derivative of the full inverse propagator at the pole.  Its real and imaginary
parts therefore describe, respectively, the field-induced changes of the
resonance mass and width.

For heavy quarkonium, this pole formulation also separates two physically
different mechanisms.  Virtual color-octet states generated by the external
field give the familiar \(E1\)-Green-function-\(E1\) contribution, whereas
open decay channels already coupled to the resonance at zero field enter
through the source dependence of their self-energy.  Together with local
two-field terms, these contributions give the complete pole response and
reduce to the standard \pNRQCD{} polarizability in the stable,
energy-independent limit.

The \(\psi(3770)\) example indicates that open-flavor dressing can produce
a sizable complex normalization of the response.  In the simple
\(P\)-wave \(D\bar D\) realization considered here, the charged--neutral
threshold splitting is only a small correction; the main effect originates
from the energy dependence generated by coupling to the open continuum.
A quantitative determination of the resonance polarizability therefore
requires the source dependence of the coupled-channel dynamics itself.

\end{document}